\documentclass[showpacs,superscriptaddress,floatfix,pra,12pt]{revtex4}
\usepackage{graphicx}
\usepackage{amsmath,amssymb}
\usepackage{bm}

\input{tcilatex}

\begin{document}

\title{Spontaneously generated atomic entanglement in free space: reinforced
by incoherent pumping}
\author{Ling Zhou}
\affiliation{School of Phys. and Optoelectronic Tech., Dalian University of Technology,
Dalian 116023, PR China}
\author{Guo Hui Yang }
\affiliation{School of Phys. and Optoelectronic Tech., Dalian University of Technology,
Dalian 116023, PR China}
\author{Anil K. Patnaik\thanks{
Current Address: Air Force Research Laboratory, Propulsion Directorate,
Wright-Patterson AFB, OH 45431 USA}}
\affiliation{Department of Physics, Texas A\&M University, College Station, Texas 77843
USA}
\affiliation{Department of Physics, Wright State University, Dayton, OH 45435 USA}

\begin{abstract}
We study spontaneously generated entanglement (SGE) between two identical
multilevel atoms in free space via vacuum-induced radiative coupling. We
show that the SGE in two-atom systems may initially increase with time but
eventually vanishes in the time scale determined by the excited state
lifetime and radiative coupling strength between the two atoms. We
demonstrate that a steady-state SGE can be established by incoherently
pumping the excited states of the two-atom system. We have shown that an
appropriate rate of incoherent pump can help producing optimal SGE. The
multilevel systems offer us more chanel to establish entanglement. The
system under consideration could be realized in a tight trap or atoms/ions
doped in a solid substrate.
\end{abstract}

\date{\today }
\maketitle


\section{\protect\bigskip Introduction}

The recent development of quantum technologies strives to resolve the quest
for the best entanglement source in quantum optical systems. Though
entanglement is observed in a variety of systems, entanglement in atomic
systems are favored as more scalable and practical systems, compared to
their ``photonic only'' counterparts due to the development of reliable
state-of-the-art technologies to control the atoms one-at-a-time \cite%
{single atom} which can be precisely scaled to many atoms \cite{atom
sorting, trapped}. Many exciting developments of entanglement sources are
involved in atomic systems such as entanglement via atom-cavity coupling %
\cite{atom-cavity}, atom-atom entanglement via cavity \cite{atom-atom cavity}%
, entanglement in trapped ions/atoms \cite{trapped, trapped1}, atomic
entanglement in an optical lattice by controlled collision \cite{optical
lattice} and also atomic entanglement via external fields \cite{external1,
external}. Recently, it has also been shown that well separated atomic
ensembles can also be entangled via coherent coupling between them \cite%
{Nature ensemble entangle}. Scully has extensively discussed entanglement in
two, three and many atoms via a single photon \cite{ScullyVolga} and has
shown that such an ensemble can produce directional spontaneous emission %
\cite{scully}.

Amongst the different atomic entanglement generation processes, an
interesting and useful category is spontaneously generated entanglement
(SGE) sources via interaction of atoms with a common bath of cavity field %
\cite{atom-atom cavity}, vacuum \cite{trapped, trapped1, atom-atom vacuum},
heat bath \cite{heat bath} or even a spin chain \cite{spin}. Usually, the
baths have very short correlation time and hence potentially cause
disentanglement \cite{3-level disentangle} and decoherence \cite{decoherence}
in an entangled system. Agarwal and Patnaik \cite{anil} have shown that
coherences in two multilevel atoms can be generated from their interaction
with a common vacuum bath via the retarded dipole-dipole (\emph{dd})
coupling when they are placed in the close proximity of each other. Effect
of such vacuum induced coherence (VIC) on the collective resonance
fluorescence is discussed in \cite{keitel geometry}. SGE is particularly
interesting from the application point of view because: practical quantum
devices are \emph{often unavoidably coupled} to the environmental bath and
hence SGE can occur naturally.

\begin{figure}[b]
\includegraphics[width=\columnwidth]{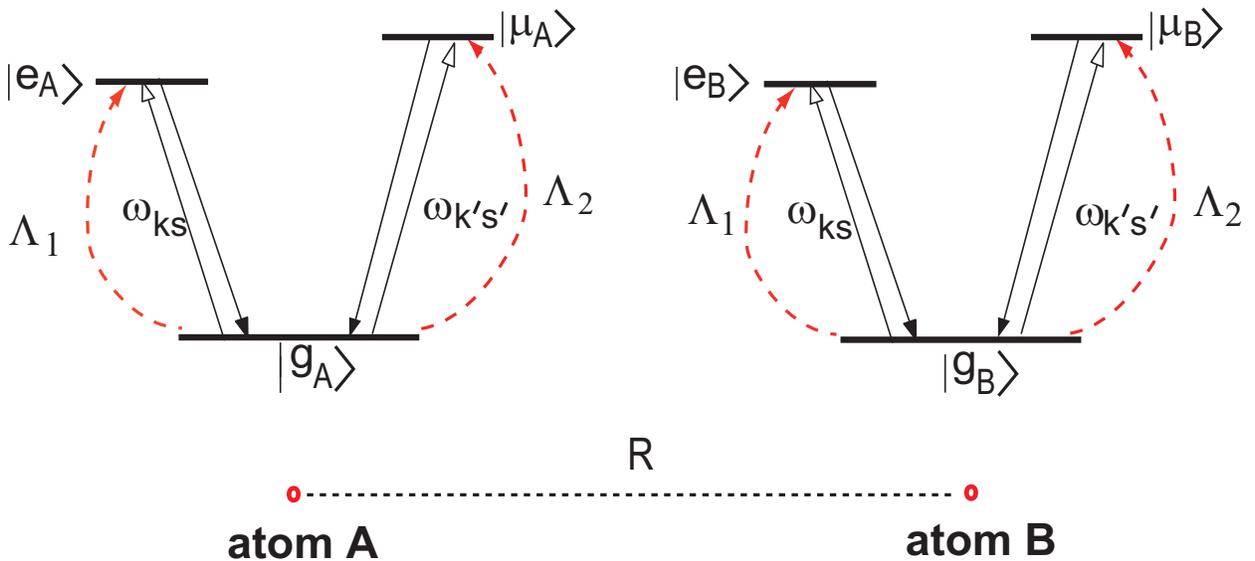}
\caption{(Color online) The two identical atoms under consideration. The V
type atoms with non-degenerate excited states. The distance between the two
atoms is considered to be small compared to the radiation wavelength, $R<%
\protect\lambda _{0}$. }
\label{Fig1}
\end{figure}

Most of the works listed above are focused on SGE in two-level atoms. Study
of multilevel systems are important because in certain situations,
participation of additional internal atomic levels in the process of
generating entanglement or causing disentanglement is unavoidable. For
example, when two atoms, having a triplet $P$-state as their excited \cite%
{anil} or ground \cite{keitel geometry} state, are placed in close
proximity, i.e., the interatomic distance is less than the wavelength of the
atomic transitions involved, $R<\lambda _{0}$, even the dipoles involving
the orthogonal $\sigma _{\pm }$ transitions can radiatively couple to
generate additional coherences. Recently, Keitel and coworkers \cite{keitel
2-lev breakdown} have explicitly shown that the two-level approximation
fails in such a situation. Furthermore, multilevel systems can open up new
channels in bath assisted SGE in a vary natural way and even can give more
control parameters \cite{keitel decoherence free}. To the best of our
knowledge, only a few studies addresses entanglement in three-level atoms
interacting with a continuum via the radiative coupling \cite{3-level
disentangle, keitel decoherence free}.

In this paper, we investigate the steady state SGE between two \emph{\
radiatively coupled and incoherently pumped atoms} having their energy
levels in a V configuration with non-degenerate excited states; see Fig. \ref%
{Fig1}. We derive a master equation and trace over the field part to obtain
the equations for the atomic dynamics. We obtain an analytical solution to
show that multilevel systems are preferable compared to two-level systems
for SGE because they add additional coupling channels to enhance the
entanglement. We demonstrate that the SGE can be sustained to achieve a
steady state entanglement by incoherently pumping the atoms. It may be noted
that we are working outside the regime of VIC. In our two-atom system VIC
could be generated if the excited states of both the atoms are degenerate or
near degenerate \cite{anil, keitel geometry}.

The organization of the paper is the following: In Sec. II we derive a
master equation for the two atoms interacting with common vacuum. Tracing
over the vacuum bath parameters, we obtain the equations for the system
dynamics. In Sec. III, we present the time evolution of the entanglement
between the two atoms that occurs only for a short period of time. In Sec.
IV, we derive the atomic density matrix equations when atoms are
incoherently pumped. We show that a steady state entanglement can be
obtained between the two atoms purely via incoherent processes. We summarize
and discuss our result in Sec. V.

\section{The two-atom system and their dynamics}

We consider two identical three level $V$ systems (say $A$ and $B$) in free
space having two excited states $|e_{\alpha }\rangle $ and $|\mu _{\alpha
}\rangle $ ($\alpha =A,B$), and a ground state $|g_{\alpha }\rangle $, as
depicted in Fig. 1. Both the atoms couple to the same vacuum field. While we
do not wish to loose the generality of our results, our scheme can
correspond to two $^{40}$Ca atoms in a magneto-optical trap (MOT) in
presence of a static magnetic field. The ground state can correspond to $%
4^{1}S_{0}$ state and the excited states can correspond to the magnetic
sublevels $4^{1}P_{1}$ of \textit{Ca} atom. The static magnetic field would
remove the degeneracy of the $4^{1}P_{1}$ sublevels and the states $%
|e_{\alpha }\rangle $ and $|\mu _{\alpha }\rangle $ can correspond to $%
m_{l}=\pm 1$ levels. Note that we restrict to the situation where the cross
couplings between $|e_{\alpha }\rangle \leftrightarrow |g_{\alpha }\rangle $
and $|\mu_{\alpha }\rangle \leftrightarrow |e_{\alpha }\rangle $ transitions
are eleminated by considering the non-degenerate excited states. Thus the
photon emitted from $|e_{\alpha }\rangle \rightarrow |g_{\alpha }\rangle $ ($%
|\mu _{\alpha }\rangle \rightarrow |g_{\alpha }\rangle $) can only be
absorbed by $|g_{\alpha }\rangle \rightarrow |e_{\alpha }\rangle $ ($%
|g_{\alpha }\rangle \rightarrow |\mu _{\alpha }\rangle $). For simplicity,
we consider only the case of real dipole moments for our discussion below.
These results can be easily generalized to complex dipoles, e.g., involving
magnetic sublevels.

In this section, we derive the system dynamics with only the contributions
from the two-atom coupling with vacuum, i.e., in absence of incoherent
pumping. The role of incoherent pumping will be discussed in detail in the
Sec. IV. The Hamiltonian of the two-atom system interacting with the vacuum
field can be written in the interaction picture as 
\begin{equation}
\begin{split}
H_{I}=& \sum_{\alpha =A,B}\Big(\sum_{ks}d_{eg}^{\alpha }\sigma _{eg}^{\alpha
}a_{ks}e^{i[\vec{k}\cdot \vec{x}_{\alpha }+(\omega _{1}-\omega _{ks})t]} \\
& +\sum_{k^{\prime }s^{\prime }}d_{\mu g}^{\alpha }\sigma _{\mu g}^{\alpha
}a_{k^{\prime }s^{\prime }}e^{i[\vec{k^{\prime }}\cdot \vec{x}_{\alpha
}+(\omega _{2}-\omega _{k^{\prime }s^{\prime }})t]}+H.c.\Big),
\end{split}
\label{eq1}
\end{equation}%
where, the vacuum Rabi coupling coefficients corresponding to atom $\alpha $
are 
\begin{eqnarray}
d_{eg}^{\alpha } &=&i\left( \frac{2\pi \hslash \omega _{ks}}{V}\right) ^{1/2}%
\vec{\wp}_{eg}^{\alpha }\cdot \vec{\varepsilon}_{ks}  \notag \\
\mathrm{and}~~d_{\mu g}^{\alpha } &=&i\left( \frac{2\pi \hslash \omega
_{k^{\prime }s^{\prime }}}{V}\right) ^{1/2}\vec{\wp}_{\mu g}^{\alpha }\cdot 
\vec{\varepsilon}_{k^{\prime }s^{\prime }},
\end{eqnarray}%
and $\vec{\wp}_{eg}^{\alpha }~(\vec{\wp}_{\mu g}^{\alpha })$ is the dipole
matrix element corresponding to the transition operator $\sigma
_{eg}^{\alpha }=|e_{\alpha }\rangle \langle g_{\alpha }|$ ($\sigma _{\mu
g}^{\alpha }=|\mu _{\alpha }\rangle \langle g_{\alpha }|$), $\vec{\varepsilon%
}_{ks}(\vec{\varepsilon}_{k^{\prime }s^{\prime }})$ is the unit polarization
vector of the vacuum mode with frequency $\omega _{ks}$ ($\omega _{k^{\prime
}s^{\prime }}$), $a_{ks}$ ($a_{k^{\prime }s^{\prime }}$) is the photon
annihilation operator corresponding to the vacuum field with wave vector $%
k~(k^{\prime })$ and polarization $s~(s^{\prime })$, $\omega _{1}$ ($\omega
_{2}$) is the atomic frequency corresponding to $|e_{\alpha }\rangle
\leftrightarrow |g_{\alpha }\rangle $ ($|\mu _{\alpha }\rangle
\leftrightarrow |g_{\alpha }\rangle $) transitions, and $x_{\alpha }$ is the
position of the atom $\alpha $.

We use the Zwanzig projection operator method \cite{mandel,agarwal} to trace
over the field degrees of freedom and obtain a reduced density matrix
equation for the atoms. We use the Born and Markoff approximation to obtain
a memoryless master equation. Referring to Ref. \cite{anil} and without
duplicating the lengthy calculation, we write the reduced density matrix
equation for the atoms as 
\begin{equation}
\dot{\rho}=-i[\mathcal{V}_{dd},\rho ]+(\mathcal{L}_{s}+\mathcal{L}_{dd})\rho
,  \label{MasterEq}
\end{equation}%
where 
\begin{equation}
\mathcal{V}_{dd}=G_{1}\sigma _{eg}^{A}\otimes \sigma _{ge}^{B}+G_{2}\sigma
_{\mu g}^{A}\otimes \sigma _{g\mu }^{B}+H.c.,
\end{equation}%
is the part of the \emph{dd}-interaction that contributes to the level
shift. The coupling coefficients are 
\begin{eqnarray}
G_{1} &=&\sum_{ks}\frac{\pi }{\hslash ^{2}(\omega _{1}-\omega _{ks})}%
d_{eg}^{A}d_{ge}^{B}e^{i\vec{k}\cdot \vec{R}},  \notag \\
G_{2} &=&\sum_{ql}\frac{\pi }{\hslash ^{2}(\omega _{2}-\omega _{ql})}d_{\mu
g}^{A}d_{g\mu }^{B}e^{i\vec{k}\cdot \vec{R}}.  \label{Gs}
\end{eqnarray}%
Here $\vec{R}=\vec{x}_{A}-\vec{x}_{B}$. Further, the Liouvillian operators $%
\mathcal{L}_{j}$ are: 
\begin{equation}
\begin{split}
\mathcal{L}_{s}\rho & =\gamma _{1}[(2\sigma _{ge}^{A}\rho \sigma
_{eg}^{A}-\sigma _{ee}^{A}\rho -\rho \sigma _{ee}^{A})+A\rightarrow B] \\
& +\gamma _{2}[(2\sigma _{g\mu }^{A}\rho \sigma _{\mu g}^{A}-\sigma _{\mu
\mu }^{A}\rho -\rho \sigma _{\mu \mu }^{A})+A\rightarrow B],
\end{split}%
\end{equation}%
corresponding to spontaneous emission of the atoms, and 
\begin{equation}
\begin{split}
\mathcal{L}_{dd}\rho & =[\Gamma _{1}(2\sigma _{ge}^{B}\rho \sigma
_{eg}^{A}-\sigma _{eg}^{A}\sigma _{ge}^{B}\rho -\rho \sigma _{eg}^{A}\sigma
_{ge}^{B})+H.c.] \\
& +[\Gamma _{2}(2\sigma _{g\mu }^{B}\rho \sigma _{\mu g}^{A}-\sigma _{\mu
g}^{A}\sigma _{g\mu }^{B}\rho -\rho \sigma _{\mu g}^{A}\sigma _{g\mu
}^{B})+H.c.],
\end{split}%
\end{equation}%
corresponding to the \emph{dd} coupling mediated by the vacuum. Note that
the subscript in $\rho _{a}$ is dropped for brevity. Here the spontaneous
decay rates are given as 
\begin{eqnarray}
\gamma _{1} &=&\frac{1}{\hbar ^{2}}\sum_{ks}\pi \delta (\omega _{1}-\omega
_{ks})|d_{eg}|^{2},  \notag \\
\gamma _{2} &=&\frac{1}{\hbar ^{2}}\sum_{ql}\pi \delta (\omega _{2}-\omega
_{ql})|d_{\mu g}|^{2},  \label{gam}
\end{eqnarray}%
and the atom-atom coupling coefficients are obtained as 
\begin{eqnarray}
\Gamma _{1} &=&\frac{1}{\hbar ^{2}}\sum_{ks}\pi \delta (\omega _{1}-\omega
_{ks})|d_{eg}|^{2}e^{i\vec{k}\cdot \vec{R}},  \notag \\
\Gamma _{2} &=&\frac{1}{\hbar ^{2}}\sum_{ks}\pi \delta (\omega _{2}-\omega
_{ql})|d_{\mu g}|^{2}e^{i\vec{k}\cdot \vec{R}}.  \label{Gamma}
\end{eqnarray}%
Further the index $\alpha $ has been dropped as we consider that the atomic
dipoles corresponding to the same atomic transitions are parallel to each
other, i.e., $\vec{\wp}_{ij}^{A}\parallel \vec{\wp}_{ij}^{B}$. Clearly, the
radiative coupling terms $\Gamma _{i}$ and $G_{i}$ have numerical
significance only in the limit $|kR|$ and $|k^{\prime }R|\leq 1$. In the
other limit, when the interatomic distance $R$ is too large, only the
spontaneous emission terms survive and the atoms behave as two independent
atoms. We refer to \cite{anil} for the detail steps of the calculation.

Now let us assume that initially, the two-atom state is $|e\mu \rangle $,
where the notation $|ij\rangle $$\equiv $$|i_{A}\rangle \otimes
|j_{B}\rangle $, $i,j=e,\mu ,g$. The nine two-atom basis states are $%
|ee\rangle ,$ $|e\mu \rangle $, $|eg\rangle $, $|\mu e\dot{\rangle},|\mu \mu
\rangle ,$ $|\mu g\rangle ,$ $|ge\rangle ,$ $|g\mu \rangle $ and $|gg\rangle 
$, and we number them $1$ through $9$ in the same order as above to simplify
the notations for the density matrix elements $\langle i_{A}j_{B}|\rho
|i_{A}^{\prime }j_{B}^{\prime }\rangle $. For examples, the density matrix
element $\langle e_{A}\mu _{B}|\rho |e_{A}\mu _{B}\rangle $ corresponding to
our initial state $|e\mu \rangle $ is represented as $\rho _{22}$ in the new
notation. The full density matrix equation involves 81 matrix elements but
for the above initial condition, many elements would be identically zero and
only 10 density matrix elements survive. We consider the geometry where
dipole matrix elements are orthogonal to each other and are real (as
discussed in Sec. III of \cite{anil}), such that the parameters $%
G_{i},\gamma _{i},\Gamma _{i}$ are real numbers seen Eq. (26) in \cite{anil}%
. In the following, we explicitely write the dynamics equations only for
those surviving density matrix elements as%
\begin{eqnarray}
\dot{\rho}_{22} &=&-2(\gamma _{1}+\gamma _{2})\rho _{22},  \notag \\
\dot{\rho}_{33} &=&-2\gamma _{1}\rho _{33}+2\gamma _{2}\rho _{22}-\Gamma
_{1}(\rho _{73}+\rho _{37})  \notag \\
&&-iG_{1}(\rho _{73}-\rho _{37}),  \notag \\
\dot{\rho}_{37} &=&-2\gamma _{1}\rho _{37}-\Gamma _{1}(\rho _{77}+\rho
_{33})-iG_{1}(\rho _{77}-\rho _{33}),  \notag \\
\dot{\rho}_{66} &=&-2\gamma _{2}\rho _{66}-\Gamma _{2}(\rho _{86}+\rho
_{68})-iG_{2}(\rho _{86}-\rho _{68}), \\
\dot{\rho}_{68} &=&-2\gamma _{2}\rho _{68}-\Gamma _{2}(\rho _{88}+\rho
_{66})-iG_{2}(\rho _{88}-\rho _{66}),  \notag \\
\dot{\rho}_{77} &=&-2\gamma _{1}\rho _{77}-\Gamma _{1}(\rho _{37}+\rho
_{73})-iG_{1}(\rho _{37}-\rho _{73}),  \notag \\
\dot{\rho}_{88} &=&-2\gamma _{2}\rho _{88}+2\gamma _{1}\rho _{22}-\Gamma
_{2}(\rho _{86}+\rho _{68})  \notag \\
&&-iG_{2}(\rho _{68}-\rho _{86}),  \notag \\
\dot{\rho}_{99} &=&2\gamma _{1}(\rho _{33}+\rho _{77})+2\gamma _{2}(\rho
_{66}+\rho _{88})  \notag \\
&&+2\Gamma _{1}(\rho _{37}+\rho _{73})+2\Gamma _{2}(\rho _{68}+\rho _{86}). 
\notag
\end{eqnarray}%
Note that the conjugate matrix elements $\rho _{73}$ and $\rho _{86}$
(conjugates of $\rho _{37}$ and $\rho _{68}$, respectively) also evolve.
Using the Laplace transform method, we solve the above coupled equations for
the density matrix elements with the initial condition $\rho _{22}=1$ to
obtain their time evolution as 
\begin{eqnarray}
\rho _{22}(t) &=&e^{-2(\gamma _{1}+\gamma _{2})t},  \notag \\
\rho _{33}(t) &=&\frac{\gamma _{2}e^{-2\gamma _{1}t}}{2}\times  \notag \\
&&\Big[\frac{\gamma _{2}\cosh (2\Gamma _{1}t)-\Gamma _{1}\sinh (2\Gamma
_{1}t)-\gamma _{2}e^{-2\gamma _{2}t}}{\gamma _{2}^{2}-\Gamma _{1}^{2}} 
\notag \\
&&+\frac{\gamma _{2}\cos (2G_{1}t)+G_{1}\sin (2G_{1}t)-\gamma
_{2}e^{-2\gamma _{2}t}}{\gamma _{2}^{2}+G_{1}^{2}}\Big],  \notag \\
\rho _{37}(t) &=&\frac{\gamma _{2}e^{-2\gamma _{1}t}}{2}\times  \notag \\
&&\Big[\frac{\Gamma _{1}e^{-2\gamma _{2}t}-\Gamma _{1}\cosh (2\Gamma
_{1}t)+\gamma _{2}\sinh (2\Gamma _{1}t)}{\Gamma _{1}^{2}-\gamma _{2}^{2}} 
\notag \\
&&-\frac{iG_{1}\cos (2G_{1}t)-i\gamma _{2}\sin (2G_{1}t)-iG_{1}e^{-2\gamma
_{2}t}}{G_{1}^{2}+\gamma _{2}^{2}}\Big], \\
\rho _{66}(t) &=&\frac{\gamma _{1}e^{-2\gamma _{2}t}}{2}\times  \notag \\
&&\Big[\frac{\gamma _{1}\cosh (2\Gamma _{2}t)-\Gamma _{2}\sinh (2\Gamma
_{2}t)-\gamma _{1}e^{-2\gamma _{1}t}}{\gamma _{1}^{2}-\Gamma _{2}^{2}} 
\notag \\
&&-\frac{\gamma _{1}\cos (2G_{2}t)+G_{2}\sin (2G_{2}t)-\gamma
_{1}e^{-2\gamma _{1}t}}{\gamma _{1}^{2}+G_{2}^{2}}\Big]  \notag \\
\rho _{68}(t) &=&\rho _{37}^{\ast }(t)\big|_{1\leftrightarrow 2},\rho
_{77}(t)=\rho _{66}(t)\big|_{1\leftrightarrow 2},  \notag \\
\rho _{88}(t) &=&\rho _{33}(t)\big|_{1\leftrightarrow 2},  \notag \\
\rho _{99}(t) &=&1-\rho _{33}(t)-\rho _{66}(t)-\rho _{77}(t)-\rho _{88}(t). 
\notag
\end{eqnarray}%
It may be noted that the initial state $\rho _{22}\equiv \langle e_{A}\mu
_{B}|\rho |e_{A}\mu _{B}\rangle $ decays with a rate of the sum of the
decays of both excited states but does not depend on the \emph{dd} coupling
terms $\Gamma _{i}$ and $G_{i}$. However, the other population and cross
terms strongly depend on the \emph{dd} coupling. The \emph{dd} terms $\Gamma
_{i}$ play the role of decays via the cosine and sine hyperbolic functions
and the coefficients $G_{i}$ cause the \emph{dd} coupling induced vacuum
Rabi oscillations. The time dependent solutions of the matrix elements show
the oscillations with frequencies determined by the atom-atom coupling
coefficients $G_{1}$ and $G_{2}$. Further, the \emph{dd} terms are strongly
dependent on the interatomic distance $R$. Hence the dynamics of the density
matrix elements are also strongly affected by $R$. We will present numerical
plots and discussions for some of the important density matrix elements that
help in evolving the SGE in the following section.

\section{Time evolution of SGE}

In this section, we calculate the time evolution of the entanglement between
two-atoms. Out of various different methods to calculate entanglement
between the two atoms, we choose the negativity, defined by \cite{Gv} 
\begin{equation}
N(\rho )=\frac{\Vert \rho ^{T_{A}}\Vert -1}{2}=-\sum_{i}\text{ }^{^{\prime
}}\lambda _{i},
\end{equation}%
as the measure of entanglement. Here $\Vert \rho ^{T_{A}}\Vert $ denotes the
trace norm of $\rho ^{T_{A}}$ \cite{Zy}; $\rho ^{T_{A}}$ is the partial
transposition matrix of the atomic system density operator $\rho (t)$. The
primed sum in the above equation represents the sum over only the negative
eigenvalues $\lambda _{i}$ of $\rho^{T_{A}}$. For a high dimensional system,
while a non-zero $N(\rho)$ is a sufficient condition to prove that a system
entangled, but a null $N(\rho)$ does not necessarily qualify a system to
become disentangled. From the definition, the negativity $N$ can also be
greater than 1. For different dimensions of the density matrix, the maximum
value of $N$ is different. For a two-atom three-level system, such as ours,
the state $\Psi =\frac{1}{\sqrt{3}}[|ee\rangle +|\mu \mu \rangle +|gg\rangle 
$ is maximally entangled one with the negativity $N(\Psi )=1$.

Thus to obtain negativity in our two-atom system, we first calculate the
eigenvalues of $\rho ^{T_{A}}$ by using the density matrix $\rho (t)$. After
a lengthy calculation, we obtain the exact eigenvalues as 
\begin{eqnarray}
\lambda _{1} &=&\lambda _{2}=0,\lambda _{3}=\rho _{22}(t),\lambda _{4}=\rho
_{33}(t),  \notag \\
\lambda _{5} &=&\rho _{66}(t),\lambda _{6}=\rho _{77}(t),\lambda _{7}=\rho
_{88}(t),  \notag \\
\lambda _{8} &=&\frac{\rho _{99}(t)}{2}+\frac{1}{2}\sqrt{\rho
_{99}^{2}(t)+4[|\rho _{37}(t)|^{2}+|\rho _{68}(t)|^{2}]}, \\
\lambda _{9} &=&\frac{\rho _{99}(t)}{2}-\frac{1}{2}\sqrt{\rho
_{99}^{2}(t)+4[|\rho _{37}(t)|^{2}+|\rho _{68}(t)|^{2}]}.  \notag
\end{eqnarray}%
It is clear that among all of the above eigenvalues, only $\lambda _{9}$ can
become negative. Therefore, if we obtain a negative $\lambda _{9}$, it is
sufficient to prove the occurence of the SGE. Clearly, $\lambda _{9}$ can be
negetive only if at least one of the matrix element $\rho _{37}\equiv
\langle eg|\rho |ge\rangle $ or $\rho _{68}\equiv \langle \mu g|\rho |g\mu
\rangle $ is non-zero, i.e., if there is an exchange of at least one photon
between two atoms. Thus it is clearly established that radiative coupling
leads to SGE in the two-atom system. Because both $\rho _{37}$ and $\rho
_{68}$ contribute to the generation of entanglement. Thus the three-level
atoms offer us more channels to establish entanglement than two-level one.

\begin{figure}[tbph]
\includegraphics[width=\columnwidth]{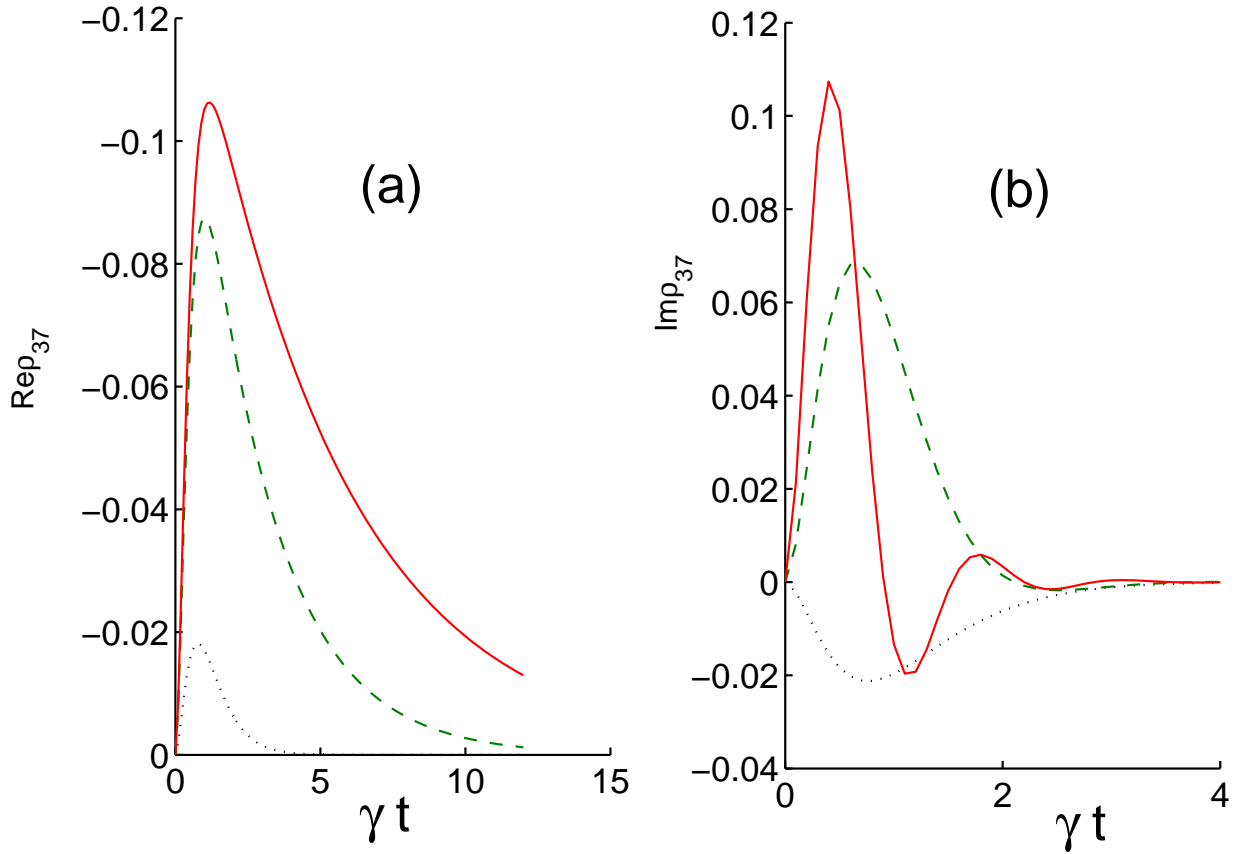}
\caption{ (Color online) The time-dependence of real and imaginary parts of
matrix element $\protect\rho _{37}(t)$ with different values of $\Gamma
_{1}(\Gamma _{2})$ and $G_{1}(G_{2})$. We set $\protect\gamma =1$,$r=1.2$.
The parameters ($G,\Gamma )\equiv (2.4,0.9)$, $(0.9,0.8)$, and ($-0.24,0.2$)
correspond to $R=0.83\protect\lambda _{1},~1.18\protect\lambda _{1}$ and $%
2.78\protect\lambda _{1}$ that are represented in the figure as the solid,
dashed and dotted line, respectively. }
\end{figure}

Before we proceed further, we first study the density matrix elements $\rho
_{37}(t)$ and $\rho _{99}(t)$ that determine the entanglement. Note that the
two excited states are non-degenerate. Assuming that $\omega _{2}=r\omega
_{1}$, and say $\gamma _{1}=\gamma $; $\Gamma _{1}=\Gamma $, $G_{1}=G$, we
have $\gamma _{2}=r\gamma $, $\Gamma _{2}=r\Gamma $ and $G_{2}=rG$. For the
numerical plots presented below, we use the parameters given in the Table %
\ref{T1} determined from their definitions in Eqs. (\ref{Gs}, \ref{gam}, \ref%
{Gamma}) \cite{numerics}. Note that as the interatomic distance reduces, $%
\Gamma$ oscillates \cite{anil}. However, peak value of $\Gamma$ decreases
with larger interatomic distances due to decreased dipole-dipole coupling
between the two atoms. Thus, $\Gamma _{i}$ ( $G_{i}$) can be same in two or
more interatomic different distances. But any particular interatomic
distance determines a particular value for the pair of $(\Gamma_i, ~G_i)$.
We will concentrate on the the photon exchange process with decreased trend
of $\Gamma _{i}$ ( $G_{i}$ ) with the increasing interatomic separation. So,
the parameters given in table is approximately monotonic. 
\begin{table}[h]
\caption{The interatomic distance and the corresponding coupling parameters.
All the frequency units are scaled with $\protect\gamma $.}
\label{T1}%
\begin{tabular}{|c|c|c|c|c|}
\hline\hline
$R$ (in unit of $\lambda _{0}$) & $\Gamma _{1}$ & $\Gamma _{2}$ & $G_{1}$ & $%
G_{2}$ \\ \hline
0.50 & 0.96$\gamma _{1}$ & 0.96$\gamma _{2}$ & 8.0$\gamma _{1}$ & 8.0$\gamma
_{2}$ \\ \hline
0.83 & 0.9$\gamma _{1}$ & 0.9$\gamma _{2}$ & 2.4$\gamma _{1}$ & 2.4$\gamma
_{2}$ \\ \hline
1.18 & 0.8$\gamma _{1}$ & 0.8$\gamma _{2}$ & 0.9$\gamma _{1}$ & 0.9$\gamma
_{2}$ \\ \hline
2.78 & 0.2$\gamma _{1}$ & 0.2$\gamma _{2}$ & -0.24$\gamma _{1}$ & -0.24$%
\gamma _{2}$ \\ \hline\hline
\end{tabular}%
\end{table}

In Fig. 2, we present the evolution of the density matrix element $\rho
_{37}=\langle eg|\rho |ge\rangle $ representing the single photon radiative
coupling process, in which atom $A$ loses its excitation to excite atom $B$
from its ground state $|g_{A}\rangle $ to the state $|e_{B}\rangle $. It is
observed that, initially the process of exchange of photon increases.
However, after reaching a maximum, Re$\rho _{37}$ falls off quickly within
one spontaneous emission cycle. The time needed to reach the maximum value
for Re$\rho _{37}$ is determined by $\Gamma ^{-1}$. The maximum of Im$\rho
_{37}$ occurs at $t\sim \gamma ^{-1}$. In long time limit both real and
imaginary part of $\rho _{37}$ vanish. Similar conclusions can be derived
for the matrix element $\rho _{68}$ which physically represents the
simultaneous probability of two processes $|\mu _{A}\rangle \rightarrow
|g_{A}\rangle $ and $|g_{B}\rangle \rightarrow |\mu _{B}\rangle $.

\begin{figure}[tbph]
\includegraphics[width=0.8\columnwidth]{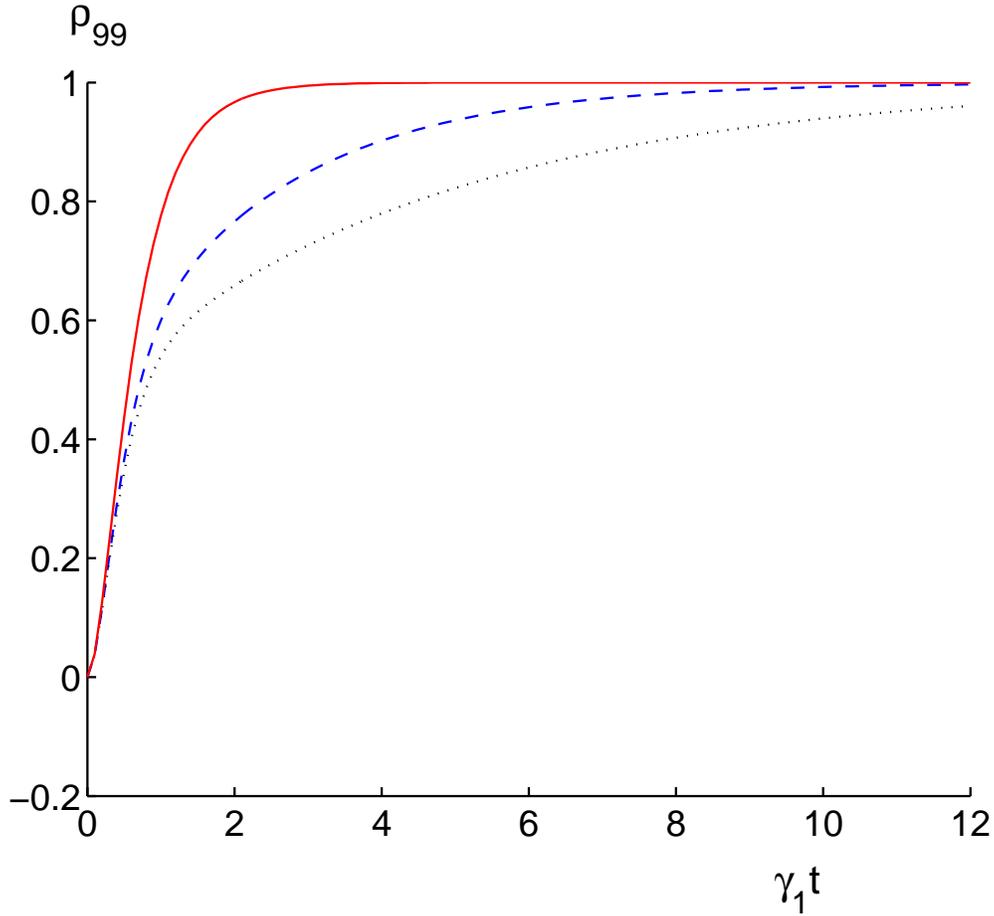}
\caption{(Color online) The matrix element for our atomic system as a
function of $t$ corresponding to different pairs of ($G,$ $\Gamma )$. The
curves and also corresponding legends are same as in Fig.2.}
\end{figure}

In Fig. 3 we present the evolution of the population in the state $%
|gg\rangle $, i.e., the matrix element $\rho _{99}(t)$. This plot also
supports the physical process we described above. We can see that again
large values of $\Gamma$ slows down $\rho _{99}(t)$ to reach at its
asymptotic value. In other words, the radiative coupling process survives
longer. The steady state value is $\rho_{99} = 1$, i.e., both the atoms
reach their ground states in the long time limit.

\begin{figure}[tbph]
\includegraphics[width=0.8\columnwidth]{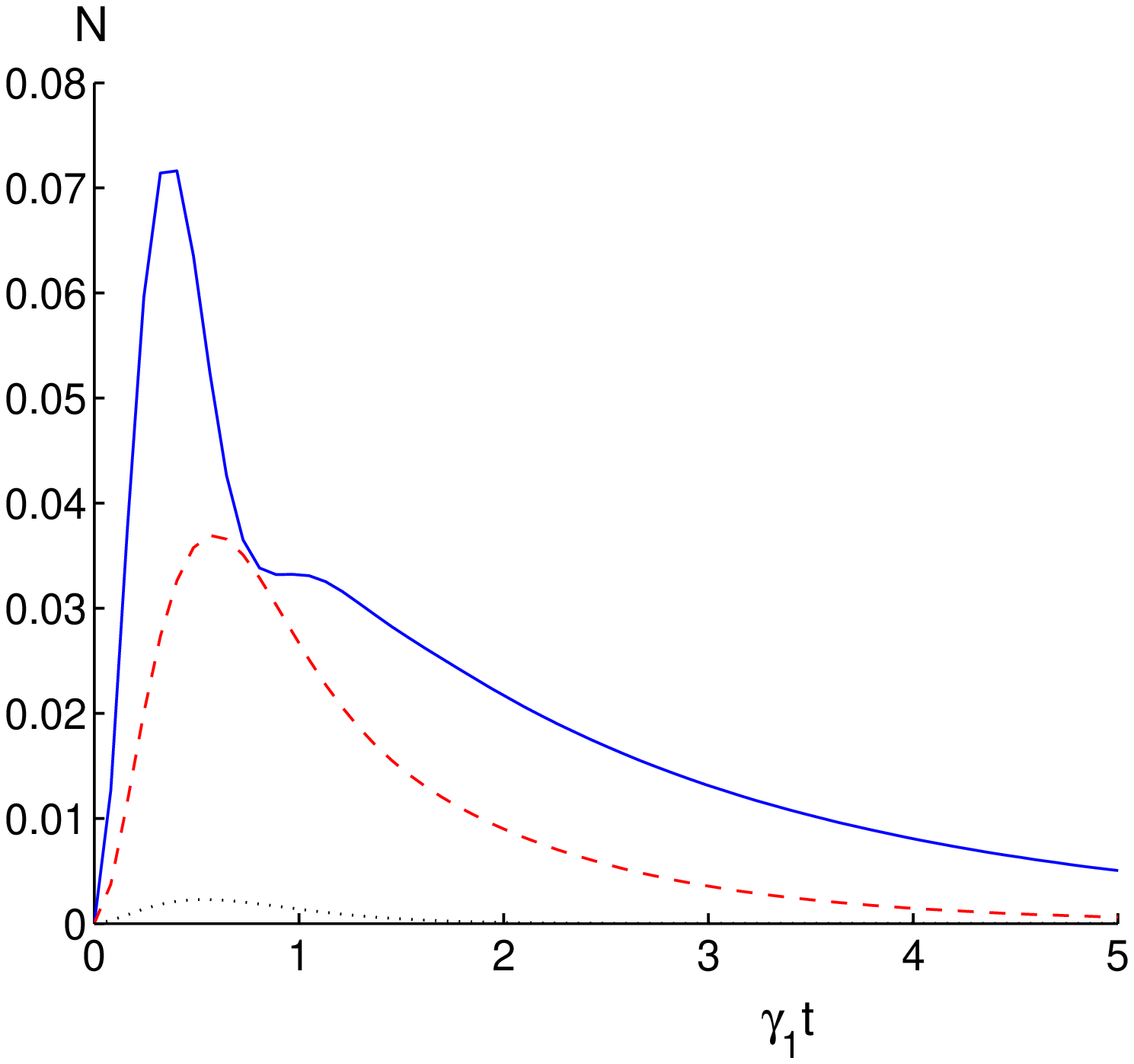}
\caption{(Color online) The negativity for our atomic system as a function
of $t$ corresponding to different pairs of ($G,$ $\Gamma )$. All the
parameters correspond to Fig.2, respectively. }
\end{figure}

Next, we discuss the property of the entanglement generated this atomic
system. We present the plot of $N(t)$ that describes the time evolution of
the SGE for different values of $\Gamma $ and $G$ in Fig. 4. It is shown
that at $t=0$, there is no entanglement because initially the atomic system
is in the state $|e\mu \rangle $. For $t>0$ the entanglement evolves to
reach its maximum value, and then undergoes a process of disentanglement.
Finally, steady state of negativity becomes identically zero. But the
relaxation time becomes longer when the atoms are nearer (seen solid line in
Fig. 2a and Fig. 4 ). From the solution Eq.(11), we know that relaxation
time of Re ($\rho _{37})$ is determined by $max\{\frac{1}{2(\gamma
_{1}+\Gamma _{1})},\frac{1}{2(\gamma _{1}-\Gamma _{1})}\}$ which means
larger the value of $|\Gamma _{1}|$, longer the relaxation time for
disentanglement. This is the competition between \emph{dd} coupling and the
all-direction spontaneous emission. The evolution of SGE presented here can
be understood as due to the following multiple radiative exchange processes:
spontaneous emission of atom A via $|e_{A}\rangle \rightarrow |g_{A}\rangle $
(atom B via $|\mu _{B}\rangle \rightarrow |g_{B}\rangle $) transition is
followed by an absorption of the spontaneously emitted radiation by atom B
via $|g_{B}\rangle \rightarrow |e_{B}\rangle $ (atom A via $|g_{A}\rangle
\rightarrow |\mu _{A}\rangle $), as it is expressed by $\rho _{37}(t)$ ($%
\rho _{68}(t)$) causing the two-atom entanglement.

\section{Steady state entanglement with broadband incoherent pumping}

From the previous section, we have seen that the disadvantage of SGE is the
temporal evolution and the quickly diminishing of SGE due to the decays in
the system. However, for any useful application, for example, to use the
two-atom system as a coupled qubit, sustaining the generated SGE is
essential. In order to achieve a steady state entanglement, we introduce an
incoherent pump to continually repump population to the excited states.
Steady state entanglement using a classical coherent pumping is discussed in %
\cite{cakir} where the coherent pumping with single frequency drive the
atoms. Coherent field assisted entanglement is rather intutive. On the
contrary, the usual notion associated with an incoherent pumping with a
decoherence process and hence a source of disentanglement. However, in what
follows below, we show that an appropriate strength of incoherent pump can
lead to a steady entanglement between the two atoms.

We consider a broadband incoherent pump acting on the two atoms which
incoherently drive the population from $|g\rangle $ to $|e\rangle $ ($|\mu
\rangle $) levels. Incoherent pumping can be modeled as an inverse process
of spontaneous emission \cite{mu,k}. Thus we add a third Liouvillian $%
\mathcal{L}_{\mathrm{inc}}$ to the the master equation Eq.(4) to get 
\begin{equation}
\dot{\rho}=-i[\mathcal{V}_{dd},\rho ]+(\mathcal{L}_{s}+\mathcal{L}_{dd}+%
\mathcal{L}_{\mathrm{inc}})\rho ,  \label{Incoh}
\end{equation}%
where 
\begin{eqnarray}
\mathcal{L}_{\mathrm{inc}}\rho &=&\Lambda _{1}[(2\sigma _{eg}^{A}\rho \sigma
_{ge}^{A}-\sigma _{gg}^{A}\rho -\rho \sigma _{gg}^{A})+A\rightarrow B] 
\notag \\
&&+\Lambda _{2}[(2\sigma _{\mu g}^{A}\rho \sigma _{g\mu }^{A}-\sigma _{\mu
\mu }^{A}\rho -\rho \sigma _{\mu \mu }^{A})+A\rightarrow B].  \notag \\
&&
\end{eqnarray}%
where $\Lambda _{1}$ and $\Lambda _{2}$ denote incoherent pumping rates for $%
|g_{\alpha }\rangle \rightarrow |e_{\alpha }\rangle $ and $|g_{\alpha
}\rangle \rightarrow |\mu _{\alpha }\rangle $ transitions respectively. We
explicitly write the density matrix equations involved in presence of the
incoherent pump as 
\begin{eqnarray}
\dot{\rho}_{11} &=&-4\gamma _{1}\rho _{11}+2\Lambda _{1}(\rho _{77}+\rho
_{33})  \notag \\
\dot{\rho}_{22} &=&-2(\gamma _{1}+\gamma _{2})\rho _{22}+2\Lambda _{1}\rho
_{88}+2\Lambda _{2}\rho _{33},  \notag \\
\dot{\rho}_{33} &=&-2s_{1}\rho _{33}+2\gamma _{2}\rho _{22}+2\gamma _{1}\rho
_{11}+2\Lambda _{1}\rho _{99}  \notag \\
&&-iG_{1}(\rho _{73}-\rho _{37})-\Gamma _{1}(\rho _{73}+\rho _{37}),  \notag
\\
\dot{\rho}_{37} &=&-2s_{1}\rho _{37}-\Gamma _{1}(\rho _{77}+\rho _{33}) 
\notag \\
&&+2\Gamma _{1}\rho _{11}-iG_{1}(\rho _{77}-\rho _{33}),  \notag \\
\dot{\rho}_{44} &=&-2(\gamma _{1}+\gamma _{2})\rho _{44}+2\Lambda _{1}\rho
_{66}+2\Lambda _{2}\rho _{77}  \notag \\
\dot{\rho}_{55} &=&-4\gamma _{2}\rho _{55}+2\Lambda _{2}(\rho _{66}+\rho
_{88}) \\
\dot{\rho}_{66} &=&-2s_{2}\rho _{66}+2\gamma _{1}\rho _{44}+2\gamma _{2}\rho
_{55}+2\Lambda _{2}\rho _{99}  \notag \\
&&-iG_{2}(\rho _{86}-\rho _{68})-\Gamma _{2}(\rho _{86}+\rho _{68}),  \notag
\\
\dot{\rho}_{68} &=&-2s_{2}\rho _{68}-\Gamma _{2}(\rho _{88}+\rho
_{66})+2\Gamma _{1}\rho _{55}-iG_{2}(\rho _{88}-\rho _{66}),  \notag \\
\dot{\rho}_{77} &=&-2s_{1}\rho _{77}+2\gamma _{1}\rho _{11}+2\gamma _{2}\rho
_{44}+2\Lambda _{1}\rho _{99}  \notag \\
&&-\Gamma _{1}(\rho _{37}+\rho _{73})-iG_{1}(\rho _{37}-\rho _{73}),  \notag
\\
\dot{\rho}_{88} &=&-2s_{2}\rho _{88}+2\gamma _{1}\rho _{22}+2\gamma _{2}\rho
_{55}+2\Lambda _{2}\rho _{99}  \notag \\
&&-\Gamma _{2}(\rho _{86}+\rho _{68})-iG_{2}(\rho _{68}-\rho _{86}),  \notag
\\
\dot{\rho}_{99} &=&2\gamma _{1}(\rho _{33}+\rho _{77})+2\gamma _{2}(\rho
_{66}+\rho _{88})-4(\Lambda _{1}+\Lambda _{2})\rho _{99}  \notag \\
&&+2\Gamma _{1}(\rho _{37}+\rho _{73})+2\Gamma _{2}(\rho _{68}+\rho _{86}). 
\notag
\end{eqnarray}%
where $s_{\alpha }=\gamma _{\alpha }+\Lambda _{1}+\Lambda _{2}$ ($\alpha
=1,2 $). Note that in presence of the incoherent pumping $\Lambda _{i}$, the
non-zero additional terms are the populations $|ee\rangle ,~|\mu \mu \rangle
,~|\mu e\rangle $. Hence we have a total of 13 non-vanishing density matrix
elements in presence of $\Lambda _{i}$. Since we are looking for a steady
state SGE, we calculate the steady state values of the matrix elements by
setting the differentials in the left-hand sides as zero and solving the
coupled equations. The analytical solutions that we obtain are 
\begin{eqnarray}
\rho _{11} &=&\frac{\gamma _{2}}{b}\Lambda _{1}a_{2},  \notag \\
\rho _{22} &=&\frac{\gamma _{1}\gamma _{2}}{b(\gamma _{1}+\gamma _{2})}%
(a_{1}\Lambda _{1}+a_{2}\Lambda _{2}),  \notag \\
\rho _{33} &=&\frac{\gamma _{1}\gamma _{2}}{b}a_{2},  \notag \\
\rho _{37} &=&\frac{\Gamma _{1}\gamma _{2}}{s_{1}b}(\Lambda _{1}-\gamma
_{1})a_{2},~~\rho _{44}=\rho _{22},  \notag \\
\rho _{55} &=&\frac{\gamma _{1}a_{1}}{b}\Lambda _{2},~~\rho _{66}=\frac{%
\gamma _{1}\gamma _{2}a_{1}}{b},  \notag \\
\rho _{68} &=&\frac{\gamma _{1}\Gamma _{2}}{s_{2}b}(\Lambda _{2}-\gamma
_{2})a_{1},  \notag \\
\rho _{77} &=&\rho _{33},~~\rho _{88}=\rho _{66},  \notag \\
\rho _{99} &=&\frac{2\gamma _{1}\gamma _{2}}{b}(\beta _{1}a_{2}+a_{1}\beta
_{2}),
\end{eqnarray}%
with 
\begin{eqnarray}
a_{1} &=&\Lambda _{2}[\gamma _{1}+2\beta _{1}(\gamma _{1}+\gamma _{2})], 
\notag \\
a_{2} &=&\Lambda _{1}[\gamma _{2}+2\beta _{2}(\gamma _{1}+\gamma _{2})], 
\notag \\
\beta _{j} &=&\frac{s_{\alpha }\gamma _{\alpha }^{2}+\Gamma _{\alpha
}^{2}(\Lambda _{\alpha }-\gamma _{\alpha })}{2s_{\alpha }\gamma _{\alpha
}(\Lambda _{1}+\Lambda _{2})},~j=1,2,  \notag \\
b &=&2\gamma _{1}\gamma _{2}(\gamma _{1}+\gamma _{2})[(\beta
_{1}+1)a_{2}+(\beta _{2}+1)a_{1}]  \notag \\
&&+(\gamma _{1}+\gamma _{2})(a_{2}\gamma _{2}\Lambda _{1}+a_{1}\gamma
_{1}\Lambda _{2})  \notag \\
&&+\gamma _{1}\gamma _{2}(2a_{2}\Lambda _{2}+2a_{1}\Lambda _{1})
\end{eqnarray}%
It is interesting to note that the steady state of the density matrix
elements do not depend on $G_{1}$ and $G_{2}$. The level shift parameters $%
G_1$ and $G_2$ typically contribute to oscillation of population and
coherence terms. Hence, in the long time limit, such fast oscillation terms
vanish. Thus steady state solutions in Eq. (17) are independent of $G_{1}$
and $G_{2}$.

Once again, as in the previous section, we calculate the eigenvalues of $%
\rho ^{T_{A}}$ to measure the entanglement. We obtain equation for the
non-zero eigen values as 
\begin{eqnarray}
&&(\rho _{11}-\lambda )(\rho _{55}-\lambda )(\rho _{99}-\lambda )-|\rho
_{37}|^{2}(\rho _{55}-\lambda )  \notag \\
&&-|\rho _{68}|^{2}(\rho _{11}-\lambda )=0.
\end{eqnarray}%
We obtain the numerical values of $\lambda $ solving the above equation and
substitute in Eq. (12) to obtain the steady state negativity as a function
of $\Lambda _{i}$, as shown in Fig. 5. We have scaled the incoherent pumping
rate $\Lambda _{i}$ with the spontaneous decay rate $\gamma _{1}=\gamma $
and also for simplicity we have assumed $\Lambda _{1}=\Lambda _{2}=\Lambda $%
. Clearly, a non-zero steady state entanglement is obtained by incoherently
repumping the excited state. Smaller the interatomic distance, larger is the
steady state entanglement. Further, as the incoherent pumping rate is
increased, the SGE increases but after reaching a certain optimal value at
around $\Lambda =0.08\gamma $, the atomic entanglement starts to reduce. For
smaller the interatomic distances, even stronger incoherent pumping can be
used to get entangled atoms. Without any incoherent pumping the steady state
SGE is identically zero. Physically, the increase in entanglement with the
incoherent pumping can be understood as follows: the spontaneous emission in
either of the two atoms followed by exchange of photon between them
generates SGE. But that does not survive longer because the spontaneously
emitted photon can escape in any arbitrary direction. Once both atoms loose
their excitation, SGE vanishes. An incoherent pump assists the atoms to
bring back to the desirable excitation so that more spontaneous emissions
and hence photon exchanges can take place between the two atoms. Thus,
increasing the repumping via incoherent pumping helps increasing the SGE.
However, incoherent repumping also competes with the photon exchange process
to re-excite the atoms. While an excitation due to the photon-exchange
process enhances the entanglement, an excitation by incoherent process has
no direct contribution to the entanglement. In fact, for a larger $\Lambda $%
, the incoherent excitation dominates the photon exchange process and hence
causes a decrease in SGE. For $\Lambda \gg \Gamma _{i},~G_{i}$, SGE becomes
identically zero.

\begin{figure}[tbph]
\includegraphics[width=\columnwidth]{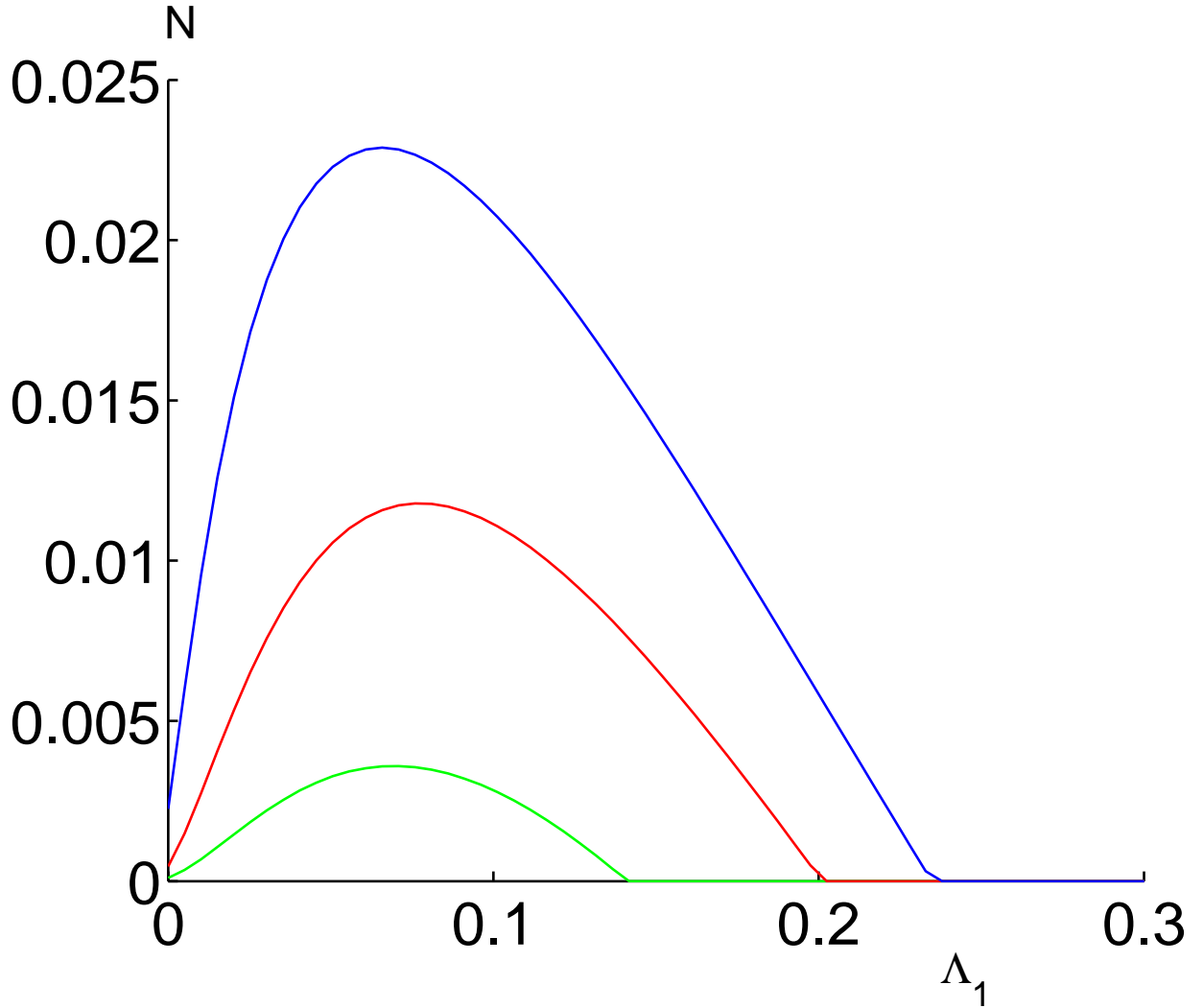}
\caption{(Color online) The steady state negativity for our atomic system as
a function of $\Lambda ,$ where we set $\Lambda _{1}=\Lambda _{2}=\Lambda ,$ 
$\protect\gamma _{1}=\protect\gamma =1$, $r=1.2$, $\protect\gamma _{2}=r%
\protect\gamma $. From bottom to top, the values of $\Gamma =0.8,0.9$ and $%
0.96$, corresponding to $R=1.18\protect\lambda _{1}$, $0.83\protect\lambda %
_{1}$ and $0.5\protect\lambda _{1}$, respectively. }
\end{figure}

\section{Discussion and Summary}

We have investigated the spontaneously generated entanglement in a system of
two three-level atoms are coupled to the common vacuum field. We have
presented the time evolution of SGE due to the photon exchange between the
two atoms. We have shown that both the magnitude of entanglement and the
survival period of SGE are enhanced by reducing the interatomic distance.
From our analytical calculations, we have shown the strong dependence of the
SGE on the radiative coupling parameters. We have explicitly demonstrated
that the multilevel atoms are preferable compared to their two-level
counterparts for SGE, because each channel adds to enhance the magnitude of
the entanglement. In the long time limit, however, SGE vanishes.

Further, to reinforce the above short term evolution of SGE in the
radiatively-coupled two-atom system, we have proposed to use an incoherent
pump that assists in repumping the deexcited atoms and sustain the SGE. We
have demonstrated that for a certain range of incoherent pumping, the steady
state value of SGE increases as it prevents atoms from loosing their excited
state population. However, since incoherent pumping competes with the
two-atom photon exchange process to reexcite the atoms, a stronger
incoherent pumping is shown to be undesirable. We have shown that an
appropriate rate of incoherent pump can help producing optimal SGE.

The above entanglement can further increase (not discussed here) if one
considers atoms having degenerate or near degenerate excited states in their
excited state that has additional coherences \cite{anil}, which will be
discussed elsewhere. The radiative coupling discussed above can be realized
in any a tight ion trap. However, this work can be generalized to realizing
SGE in a chain of quantum dots or even in a typical dense multi-atom system.
We believe this work will open up a new way to utilize the naturally
occurring SGE to realize an efficient entanglement source.

This work was supported by NSFC under Grant No.10774020, and also supported
by SRF for ROCS, SEM. AKP is indebted to Prof. G. S. Agarwal, Dr. P.
Anisomov, and Prof. M. O. Scully for discussions on various aspects of the
two-atom multilevel systems.


\begin{thebibliography}{99}
\bibitem{numerics} The numerical parameters are obtained for a chosen
geometry with $\theta =\frac{\pi }{2}$, $\phi =\frac{\pi }{4}$ in Eq. (26)
of \cite{anil}.
\end{thebibliography}
\end{document}